\def\full{} %

 \PassOptionsToClass{hyphens}{url}

\ifdefined\full
\documentclass[sigconf,nonacm,urlbreakonhyphens]{acmart}
\else
\RequirePackage{pdf14}
\documentclass[sigconf,urlbreakonhyphens]{acmart}

\copyrightyear{2022}
\acmYear{2022}
\setcopyright{rightsretained}
\acmConference[ARES 2022]{The 17th International Conference on Availability, Reliability and Security}{August 23--26, 2022}{Vienna, Austria}
\acmBooktitle{The 17th International Conference on Availability, Reliability and Security (ARES 2022), August 23--26, 2022, Vienna, Austria}\acmDOI{10.1145/3538969.3544445}
\acmISBN{978-1-4503-9670-7/22/08}
\fi

\usepackage{cleveref}

\usepackage{amsmath}
\usepackage{rotwein}
\usepackage{textcomp}

\newcommand{\rom}[1]{(\uppercase\expandafter{\romannumeral #1\relax})}

\newcommand{\mypar}[1]{\textbf{#1}:}

\title[YOU SHALL NOT COMPUTE on my Data: \\Access Policies for Privacy-Preserving Data Marketplaces \\and an Implementation for a Distributed Market using MPC]{Y{\huge OU} S{\huge HALL} N{\huge OT} C{\huge OMPUTE} on my Data: \\\LARGE Access Policies for Privacy-Preserving Data Marketplaces \\and an Implementation for a Distributed Market using MPC}

\ifdefined\full
\titlenote{This paper was published in the 17th International Conference on Availability, Reliability and Security (ARES 2022), August 23--26, 2022, Vienna, Austria, ACM. \url{https://doi.org/10.1145/3538969.3544445}}
\fi

\author{Stefan More}
\orcid{0000-0001-7076-7563}
\email{stefan.more@iaik.tugraz.at}
\affiliation{%
  \institution{Graz University of Technology}
  \city{Graz}
  \country{Austria}
}
\author{Lukas Alber}
\email{lukas.alber@iaik.tugraz.at}
\affiliation{%
  \institution{Graz University of Technology}
  \city{Graz}
  \country{Austria}
}


\begin{document}

\ifdefined\full\else
\fancyhead{}
\fi


\begin{abstract}
Personal data is an attractive source of insights for a diverse field of research and business.
While our data is highly valuable, it is often privacy-sensitive.
Thus, regulations like the GDPR restrict what data can be legally published, and what a buyer may do with this sensitive data.
While personal data must be protected, we can still sell some insights gathered from our data that do not hurt our privacy.
A data marketplace is a platform that helps users to sell their data while assisting buyers in discovering relevant datasets.
The major challenge such a marketplace faces is balancing between offering valuable insights into data while preserving privacy requirements.
Private data marketplaces try to solve this challenge by offering privacy-preserving computations on personal data.
Such computations allow for calculating statistics or training machine learning models on personal data without accessing the data in plain.
However, the user selling the data cannot restrict who can buy or what type of computation the data is allowed.

We close the latter gap by proposing a flexible access control architecture for private data marketplaces, which can be applied to existing data markets.
Our architecture enables data sellers to define detailed policies restricting who can buy their data.
Furthermore, a seller can control what computation a specific buyer can purchase on the data, and make constraints on its parameters to mitigate privacy breaches.
The data market's computation system then enforces the policies before initiating a computation.

To demonstrate the feasibility of our approach, we provide an implementation for the KRAKEN marketplace, a distributed data market using MPC\@.
We show that our approach is practical since it introduces a negligible performance overhead and is secure against several adversaries.
\end{abstract}

\maketitle

\section{Introduction}
\label{sec:introduction}

Due to the ongoing digitalization of society, smart appliances play an increasingly important role in our daily life.
Such smart devices continuously collect data through their sensors while providing valuable insights about us and everything surrounding us.
Personal data, in particular, has become an attractive source for insights for the individual as well as for various companies and institutions.
From regular smartwatches, to smart functional clothing for professionals tracking body metrics during training, and even invasive monitoring of vital functions in the hospital --- smart devices capture large amounts of data.
Those data sets can then be exploited using computations like traditional algorithms and novel machine learning-based approaches.
The results of such computations have proven to be valuable for different business and research fields such as medicine, marketing, and more.

In order to enhance the exploitation of such data sets, available data must be efficiently brokered to relevant consumers.
Data marketplaces take on this brokerage task.
They are an online platform that brings together the producers of personal data with relevant consumers.
However, the collected personal data is highly sensitive, and legislators tend to protect it well.
An example is the EU's General Data Protection Regulation (GDPR)~\cite{GDPR}, which defines the circumstances under which collecting, transmitting, storing, or processing such data is allowed.
Since data markets aggregate and sell valuable personal data to other companies, they must comply with regulations like the GDPR\@.
Especially data sets that might identify certain persons present a unique challenge since misuse can lead to discrimination (e.g., insurance, job market, etc.).

\textit{Private} data marketplaces~\cite{DBLP:conf/icdcs/KoutsosPCT020} try to mitigate these issues by
using modern privacy-enhancing technologies.
These technologies enable computation on personal data without revealing the data itself. Recently multiple approaches relying on this principle have been published:
For example, Enveil~\cite{enveil} uses homomorphic encryption (FHE) to provide computations on encrypted data, enabling the outsourcing of computations without revealing the data.
Wibson~\cite{DBLP:journals/corr/abs-2001-08832} is a smart contract-based marketplace that offers privacy for the identities of buyers and sellers.
Another example for a private data marketplace is Agora~\cite{DBLP:conf/icdcs/KoutsosPCT020}, which uses functional encryption (FE) to provide such privacy-preserving computations.
Agora provides verifiable output but reveals the result of a computation to the marketplace.
In contrast, KRAKEN~\cite{DBLP:conf/primelife/KochKPR20} is a marketplace architecture that also encrypts the computation's result to hide it from a curious marketplace, closing the gap left open by Agora.
The KRAKEN marketplace uses multi-party computation (MPC) to preserve users' privacy. The data is distributed in opaque shares to several nodes for computation. Only the final assembly of all the output shares discloses the result to the computation buyer.

\mypar{Challenge: Control computations on personal data}
A challenge private data marketplaces face is that users have limited ability to control what buyers can do with their data.
Further, users need to trust the marketplace to follow the rules they specify for their data.
Therefore, they need to trust that the marketplace is not covertly performing computations on their data.

In the Agora and KRAKEN marketplace, data providers cannot easily control who can buy computations on their data or access the results since the marketplace's computation system has no information about the buyer's identity.

When using functional encryption~\cite{DBLP:conf/icdcs/KoutsosPCT020}, the result of a computation is revealed to the marketplace.
On the other hand, in KRAKEN the distributed computation system encrypts the results. However, as the marketplace provides the encryption key, the computation system needs to trust the marketplace.
Thus, a curious marketplace can gain access to the computation results.
Another design variant checks the buyer's eligibility using a centralized component~\cite{KRAKEN5.4}, which is contradictory for a distributed platform.

A further challenge is restricting the type of computation a buyer can execute.
While some systems~\cite{DBLP:conf/primelife/KochKPR20} enable the user to restrict the allowed computation, and others~\cite{DBLP:conf/icdcs/KoutsosPCT020} even enforce this cryptographically, no expressive way of defining such restrictions exists yet.
For example, a seller cannot limit specific computation functions or parameters to specific buyers in existing systems.

\subsection{Contribution}

In this paper, we solve the described challenge by adding a policy system to private data marketplaces.
The contributions of our paper are as follows:

\rom{1} \textbf{Marketplace Policy Architecture:}
We introduce an architecture for an extension of private data marketplaces.
This extension features a flexible access control mechanism based on a policy system.
In the resulting marketplace system, data sellers can define expressive policies to control the usage of their data.
Those policies are attached to data products offered on a marketplace.
When a buyer purchases a computation on some data products, they are asked to provide a set of credentials certifying their identity and other attributes.
The marketplace's computation system then verifies if the credentials fulfill the policy for the selected data.
Additionally, the system uses the policy to check if the (now-authenticated) buyer is qualified to execute the requested computation.
Only then the system proceeds and executes the computation.
Further, the buyer's credentials are used to encrypt the result of a computation.
This ensures that only the legitimate buyer can access the result.

We present a generic design, so the system can be applied to various existing marketplaces.

\rom{2} \textbf{Implementation:}
To show the feasibility of our design, we provide an implementation for a distributed marketplace using multi-party computation (MPC).
In specific, we extend the architecture of the \textit{KRAKEN} marketplace~\cite{DBLP:conf/primelife/KochKPR20}.
To realize the policies, we use the \textit{TPL} system introduced by \citet{DBLP:conf/ifiptm/ModersheimSWMA19}.
By integrating the TPL system into the KRAKEN marketplace architecture, we enable users to formulate policies either in a Prolog-like programming language, or using a graphical tool.
We extend the computation nodes of KRAKEN with a interpreter for TPL policies.
The nodes then use the interpreter to check the buyer's identity and their request, before executing a computation.

In our implementation, sellers can limit what types of computation a buyer can purchase, and restrict specific computations to specific buyers.
The latter is done by first authenticating the buyer using their credentials.
To support this, we show how user's can use the features of TPL to verify the identity of data buyers using existing trust schemes like the EU's eIDAS\@.
Additionally, we apply an extension to TPL~\cite{DBLP:conf/openidentity/AlberMMS21}, which supports sellers in checking the authenticity of (verifiable) credentials~\cite{w3cVC} using distributed ledgers.
To ensure that a computation can not reveal the plaintext data of a seller, policies can also contain rules on the number of data products.

\rom{3} \textbf{Discussion and Evaluation:}
Finally, we evaluate our implementation by examining its security properties and conducting a benchmark analysis.
The benchmarks show that our implementation introduces a performance overhead of around two seconds.
Given the data-analysis computations using MPC can take from minutes up to several hours~\cite{KRAKEN4.3}, we consider this overhead acceptable.
We also discuss the security properties of our architecture by evaluating potential adversaries and attacks, and show how our approach mitigates those.
This demonstrates the practicality of our approach.

\textit{The rest of the paper is structured as follows:}
In \Cref{sec:background}, we recapitulate basic information on private data marketplaces, and on the TPL policy system we use.
Afterwards, in \Cref{sec:design}, we introduce the architecture of a private data marketplace in detail, and describe our modifications.
We also discuss in detail the process of a data trade using our modified marketplace.
Following in \cref{sec:implementation}, we describe the instantiation of our architecture using the KRAKEN marketplace and TPL, and give an example data seller policy.
Finally, in \Cref{sec:discussion}, we describe the results of our performance benchmarks, and discuss the security properties of our approach.

\section{Background}
\label{sec:background}

\subsection{Private Data Marketplaces}
\label{sec:background:datamarket}

Data Marketplaces are online platforms that enable the brokerage of data, in many cases, personal data.\footnote{e.g., aws.amazon.com/data-exchange, datacoup.com, datasift.com}
On more traditional data marketplaces, a user, or a device acting on behalf of a user, uploads personal data records to the platform, where it is arranged and categorized into sets.
Entities interested in the data can browse the marketplace's catalog and buy some data sets.
The marketplace, meanwhile, takes care of storage logistics, brokerage, and invoicing.
By providing these services, the marketplace and its operators have plaintext access to the data since it is (by design) neither encrypted nor anonymized.

In contrast, \textit{Private} Data Marketplaces enable the brokerage of personal data in a privacy-preserving way.
A user encrypts their data before uploading their data to the marketplace, hiding it from the marketplace operator and other parties who have not acquired legitimate access.
Further, advanced private marketplaces do not sell direct access to the plain data but offer (the results of) computations on the data.
The service includes combining data sets of multiple users and from various sources.
Since most data applications boil down to statistical or machine learning algorithms, this limitation does not hamper data buyers. However, it ensures compliance with data protection regulations.

Marketplaces achieve the aforementioned privacy feature by applying various cryptographic techniques, such as functional encryption (FE)~\cite{DBLP:conf/icdcs/KoutsosPCT020,DBLP:conf/edbt/Morley-Fletcher17}, fully homomorphic encryption (FHE)~\cite{enveil,DBLP:journals/tkde/NiuZWGC19}, differential-privacy (DP)~\cite{DBLP:conf/kdd/NiuZWTGC18}, or secure multi-party computation (MPC)~\cite{DBLP:conf/primelife/KochKPR20}.
Also, some marketplaces utilize Distributed Ledgers (DLs) and smart contracts to achieve additional properties like buyer or seller anonymity~\cite{DBLP:journals/corr/abs-2001-08832}.
Depending on the applied technique, the architecture either involves a centralized processing node (e.g,. FE), or a network of nodes (MPC).

\subsection{Trust Policy Language (TPL)}
\label{sec:background:tpl}

TPL ist a trust policy system introduced in 2019 by \citet{DBLP:conf/ifiptm/ModersheimSWMA19}.
The original purpose of TPL was to run an automated decision process to check whether an incoming transaction request can be trusted.
The requirements for trustworthiness in this request can be specific to each service provider, e.g., signatures with a certificate chain to some trusted entity.
TPL was initially created for the LIGHTest project and it's global trust scheme verification system~\cite{DBLP:conf/openidentity/BrueggerL16, DBLP:conf/openidentity/Rossnagel17, DBLP:conf/openidentity/OmololaMFWA19}.
Later \citet{DBLP:conf/openidentity/AlberMMS21} decoupled it, and enhanced it with support for self-sovereign identity (SSI) concepts like verifiable credentials (VCs)~\cite{w3cVC} and decentralized identifiers (DIDs)~\cite{w3cDID}.

The syntax of TPL is similar to Prolog~\cite{ISOprolog}.
A TPL policy is a set of horn clauses, each with the form $p(t)$~:-~$q_1(u_1),\dots,q_n(u_n).$
It translates to: $p(t)$ is true if all the $q_i(u_i)$ are true.
Note that we call such horn clauses rules, and a set of rules for the same $p$ defines the whole \textit{predicate}.

To run a policy check, a query $p(s)$ is evaluated and returns true if it finds suitable rules to substitute.
In more detail, a query $p(s)$ and a predicate $p(t)$ match if $s$ and $t$ can be unified.
The unifier is then applied to all $q_i(u_i)$, and each of them is evaluated.
If all subqueries are true, the query returns true.
If a subquery returns false, another rule that defines the predicate is evaluated.

Since subqueries themselves can have subqueries, the evaluation is solved by recursion until it hits a leave (i.e., a relational operation, etc.).
Only if a subquery is a so-called \textit{built-in predicate}, a callback to the backend system is executed (e.g., for server lookups, etc.).
On the success of the built-in predicate, true is returned.
Such built-in predicates are used to navigate in certain files using the concept of \textit{formats}.
They are also used for the discovery and verification of trust information, e.g., eIDAS certificates~\cite{DBLP:conf/openidentity/WagnerWMH19} or SSI credentials~\cite{DBLP:conf/trustcom/AbrahamKMRS21,DBLP:conf/openidentity/AlberMMS21}.

TPL policies can be created in the mentioned programming-language~\cite{ISOprolog}, or using a graphical authoring tool~\cite{DBLP:conf/openidentity/ModersheimN19,DBLP:conf/openidentity/WeinhardtP19,DBLP:conf/icissp/WeinhardtO19} which enables that also non-technical domain-experts create policies.

\begin{figure*}[ht!]
  \centering
\includegraphics[width=2\columnwidth]{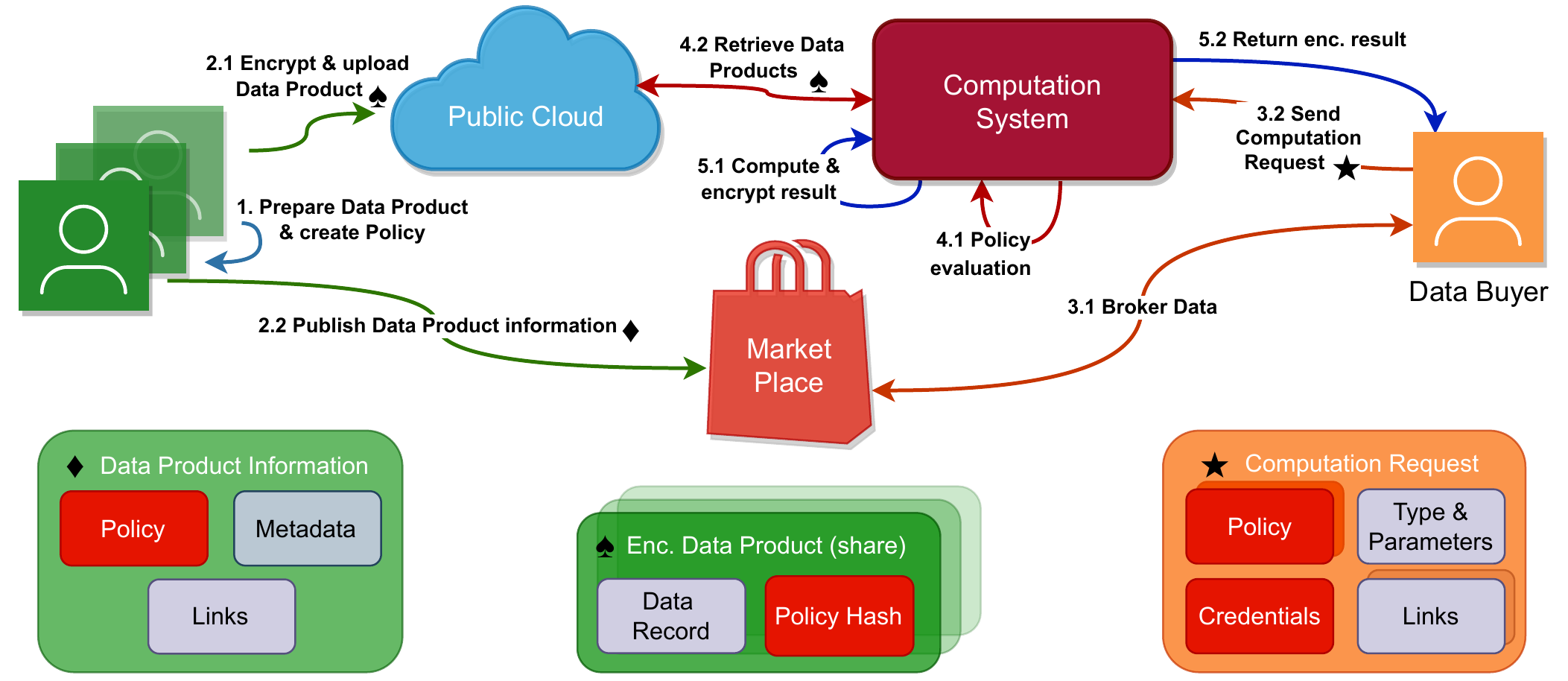}
\caption[]{Architecture and dataflows of a private data marketplace extended with our policy system. In addition to several modifications to the brokerage process (cf.~\Cref{sec:phases}), we add step 1 to create and step 4.1 to verify a policy. We also add the policy-related data, which is highlighted in red. The spade, diamond and star symbols show where the data packages are used in the flow.}\label{fig:arch}
\end{figure*}

\section{Design}
\label{sec:design}

In this section, we describe the generic design of our approach.
First, we give an overview of the components of a private data marketplace, and list the components which we add to this architecture.
Next, we explain the generic process of selling and buying data using a system which applied our modifications.
Using this extended marketplace architecture, data sellers can define the rules about who is allowed to acquire their data, and also limit which computations a buyer can perform on the data.

\textbf{Scope:}
Our design focuses on \textit{private} data marketplaces which allow a computation on user's data \textit{without the user's involvement} (non-interactive).
Thus, we don't consider systems where the user participates in computation on their data (which does not require this type of policy system).

\subsection{Architecture Overview}
\label{sec:arch}

In general, a private data marketplace (cf.~\Cref{sec:background:datamarket}) consists of the following components:

\mypar{Data Seller}
The actor who produces data and wants to offer it on the marketplace.
To host this data, the data seller uses some \textbf{public cloud storage}.
Some models subdivide the data seller further into separate roles, i.e., the data producer/generator, the data subject, and the data provider.

\mypar{Data Buyer}
The actor that wants to buy computations on the data of several data sellers.
They select one or multiple data products on a marketplace and decide which computations to execute.
The data buyer is sometimes refered to as data consumer.

\mypar{Marketplace}
The online platform which acts as a broker to connect data sellers with relevant data buyers and enables the data trade.
The marketplace provides a \textbf{catalog of data products} to which a data seller can add their \textbf{data records}.
In addition, the marketplace helps the data buyer to find data products of their liking and sells the utilization of the data on its computation infrastructure.
Additional tasks the marketplace offers are out of the scope of this paper, e.g., payment processing.

Private data marketplaces use a privacy-preserving \textbf{computation system} to perform the computation requested by a data buyer.
The number of \textbf{computation nodes} $N$ involved in this computation depends on the cryptographic technique applied by a marketplace.
In Multi-Party Computation (MPC), the computation is performed distributed on several nodes ($N > 1$), and each node only receives a part of the user's data (e.g.,~\cite{DBLP:conf/primelife/KochKPR20}).
Techniques like Functional Encryption (FE) and full homomorphic encryption (FHE) are performed on a single node ($N = 1$, e.g.,~\cite{enveil,DBLP:conf/icdcs/KoutsosPCT020}).

\vspace{1em}
In addition, our approach introduces the following additional components:

\mypar{Policy Interpreter}
The marketplace uses the policy interpreter software component to decide if a particular buyer is qualified to acquire (a computation on) some data records.
As an input the interpreter takes a \textbf{trust policy} (cf.~\Cref{sec:background:tpl}) defined by the seller for their data, as well as a set of \textbf{credentials} from the buyer, alongside some metadata about the requested computation.

\subsection{Phases}
\label{sec:phases}

In this section, we describe a private data marketplace to which we added our policy interpreter component.
We also add a step necessary to create a policy, and adapt the brokerage logic to inform users what credentials they need to provide.
A graphical overview of this extended marketplace architecture is shown in \Cref{fig:arch}.
We split the flow into the following phases:

\begin{enumerate*}
   \setcounter{enumi}{-1}
    \item To trade their data, the seller first creates an account at the marketplace.
    Additionally, they receive the cryptographic material required to sell data.
    \item The seller then prepares the data they want to sell.
    In our approach, a seller also defines the policy for their data.
    This policy specifies who can buy the data and what types of computations the seller allows.
    \item After encrypting and uploading the data to a public cloud, the seller registers the records together with the policy on the marketplace.
    They do so by combining the web links to the (encrypted) data and the policy with (unencrypted) metadata describing the data.
    Publishing this record on the marketplace creates a so-called data product,
    \item which the buyer discovers using the marketplace catalog.
    The buyer then selects a set of data products from the catalog, and specifies which computation they want to perform on those products.
    In our approach, each of the data products comes with their own policy, so there is now a set of policies which the buyer needs to fulfill.
    Before purchasing a computation, the buyer provides the required credentials to prove access qualifications w.r.t. the involved policies.
    The marketplace collects both the computation specification and the buyer's credentials.
    It then sends it alongside the selected products and policies to the computation system.
    \item A computation system with our policy extension uses the policies and the credentials to determine whether the buyer is eligible.
    \item On granted access, the system fetches the data from the clouds and performs the specified computations.
     After completing the computation, the system encrypts the data and returns the result to the buyer.
\end{enumerate*}

In the following paragraphs, we describe every single phase in more detail.
We give a concrete instantiation of this process in \Cref{sec:implementation}. %

\mypar{(0) Setup}
As a first step, users who want to become data sellers or data buyers create an account on the online marketplace.
Setup steps depend on the concrete marketplace, but usually also involve the establishment of a payment channel.
As a result of this phase, a new user
obtains cryptographic material enabling them to create data products for the marketplace.
A seller can also retrieve parts of this cryptographic material directly from a computation system they trust.
Additionally, the user receives some credentials which they can use to reauthenticate at the marketplace later (e.g., username and password, or a verifiable credential~\cite{w3cVC}).

\mypar{(1) Data and Policy Preparation}
To provide some data on the marketplace, the data seller %
first retrieves some data they want to sell, e.g., from their local storage system or IoT devices.

As additional preparation step before uploading the data, the seller formulates their policy.
Primarily, this policy contains the rules about who is eligible to buy computations on the corresponding data.
While providing a list of qualified buyers is the simplest option, it is not practical for a large set of potential buyers.
Thus, the seller could instead restrict access to a category of qualified buyers.
For example, they can require that the buyer provides a qualified certificate from a specific trust scheme (e.g., the European Union's eIDAS).
In another example, the seller may restrict the type of buyer (e.g., public universities or certified medical research organizations).
Further, the policy also contains the types of computations a particular buyer category is allowed to perform on the data.
The seller can allow different computations for different sellers.
As an alternative to formulating their own policy, the seller can browse the marketplace for existing policies and select one that suits their requirements.

\mypar{(2) Selling}
The seller then prepares the data package for selling on the marketplace, i.e., by preparing the data and encrypting it using the cryptographic material retrieved in the setup phase.
The details of this step depend on the cryptographic technique used by the specific marketplace and, thus, on the number of computation nodes $N$.
For example, for distributed computation architectures ($N > 1$), the seller first splits the data into $N$ shares.
The result of this step is a data package prepared for the respective privacy-preserving computation technique.
To restrict who can perform computations on the data, the prepared package is additionally encrypted for the specific computation node(s).

Additionally, to prevent an attacker from replacing the policy (cf.~\Cref{sec:discussion}) with their own, the seller cryptographically links the policy to the data.
They do so by adding a hash digest of the policy to the encrypted data package.

Afterward, the seller uploads the encrypted data package to a server, e.g., to a public cloud.
Then, they create a data product on the marketplace by registering the web links to the uploaded data package, alongside the policy and some metadata describing the data.

\mypar{(3) Buying}
A buyer browses the marketplace's product catalog and selects one or several data products they want to use (the selection is platform dependent and can be visual or programmatically).
Additionally, they specify the computation they want to execute on the data and initialize a computation request.

The marketplace first loads the policies of all selected data products and does an (informal) pre-check if the buyer's computation request is possible.
The system aborts at this place if the requested computation is not possible on this data.
It then computes the list of credentials required to fulfill all involved policies and sends the list to the buyer.
The buyer completes the computation request by providing all the requested credentials to the marketplace, which calls the computation system.
As an alternative, the buyer discovers the computation system using the marketplace, and forwards their credentials directly to the computation node(s).

Next, the computation system initiates the computation at the computation node(s).
Depending on the number of computation nodes $N$, the computation process on the node(s) look different.
For the special case of architectures with a single computation node ($N = 1$), only this single nodes performs the computation.
For distributed computation architectures ($N > 1$), the marketplace sends the computation request to all nodes.
All nodes then perform the same operations, but use their own key material and individual part of the input.

\mypar{(4) Policy Evaluation}
After receiving a computation request, the computation node first uses the provided links to download all data packages from the public clouds, and decrypts them.

Before the system launches any computation on the data (shares), it checks whether the buyer is entitled to the requested computation.
For this it verifies the buyer's credentials using the policies, which it receives for each data product.
Before evaluating a policy, the system checks if each policy really belongs to their data.
This check is done by calculating the hash of each policy, and comparing it with the policy hash inside the corresponding (now-decrypted) data package.

If this precheck is successful, the node launches the policy interpreter.
The inputs to the interpreter are the computation request and all the credentials the node received.
Further, the node also provides the number of retrieved data records to the interpreter.
As a first step, the interpreter checks if all involved credentials are linked to the same entity, i.e., by checking if they reference the same subject identifier~\cite{w3cDID}.
This prevents that an entity can mix the credentials of several unrelated people to fulfill the policies.

The interpreter then checks if the buyer fulfills the requirements stated by all involved sellers and if the buyer is permitted to perform the requested computation.
As part of this check, the interpreter can download additional trust status information.
For example, to verify a certificate about the buyer's legal identity, the interpreter can retrieve eIDAS trust status lists.
Another example involves SSI credentials: the interpreter can retrieve trust data registries which are stored in distributed ledgers, e.g., smart contract-based trust registries~\cite{DBLP:conf/sec/MoreGHAK21,DBLP:conf/openidentity/KubachR21}.
During these verifications, the interpreter also assumes the task of validating the revocation status of the trust data.

\mypar{(5) Computation}
If the interpreter concludes that all rules are fulfilled, the node(s) proceeds with executing the requested computation.
After the successful computation, all nodes encrypt the computation result using the buyer's public key.
Since the node extracts the public key from the buyer's primary credential, no one but the buyer can view the result.

Finally, the buyer receives the encrypted result and decrypts it using their private key.
For distributed systems with $N > 1$, the buyer receives only a part of the result from each node and has to assemble them into the final result. %

\begin{figure}[ht]
    \centering
    \includegraphics[width=\columnwidth]{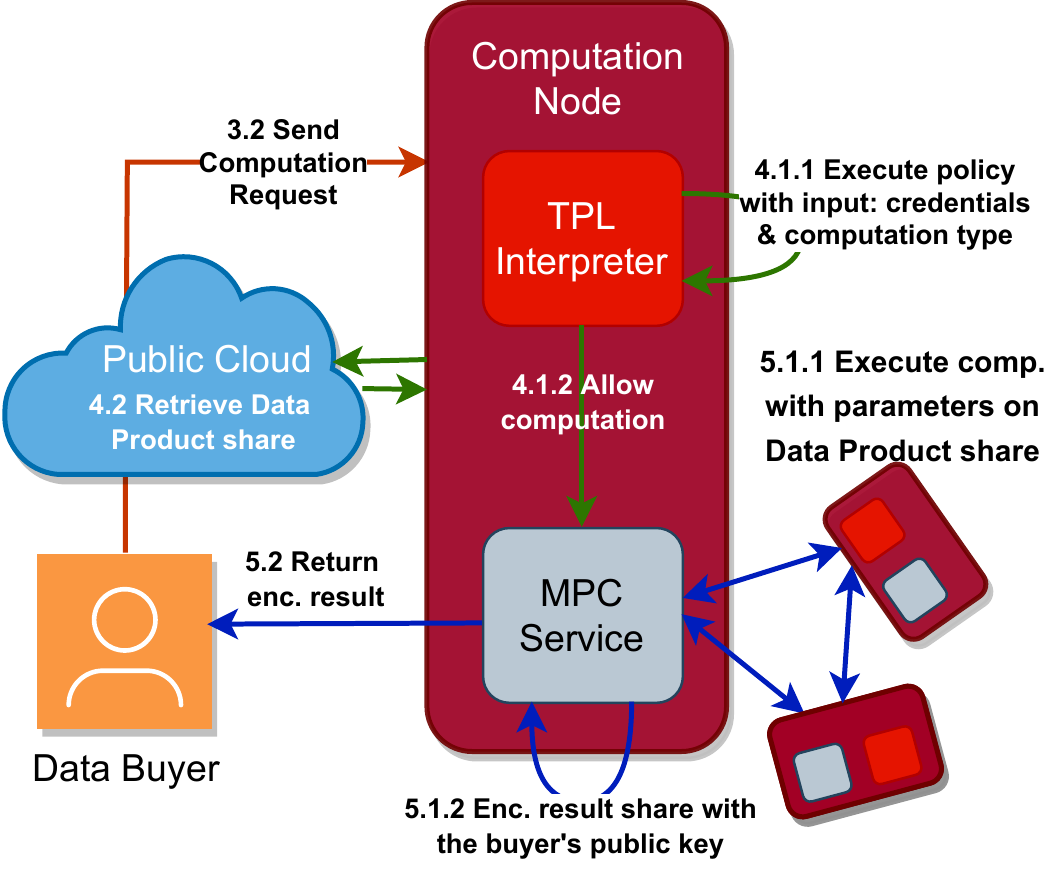}
    \caption[]{\label{fig:impl} The dataflow of a MPC computation node (cf.~\Cref{sec:implementation}), extended with our policy interpreter component. After receiving a computation request and retrieving the encrypted data products, the node first checks the validity of the computation request w.r.t. the given policies and buyer's credentials. Only on success it initiates the MPC computation.}
\end{figure}

\section{Implementation}
\label{sec:implementation}

To show the feasibility of our design, we provide a prototype implementation for the KRAKEN marketplace~\cite{DBLP:conf/primelife/KochKPR20}.
In this section, we discuss the implications of our approach on the distributed KRAKEN architecture and describe how we apply the policy component (cf.~\Cref{fig:impl}).

\mypar{Computation System}
For computations over encrypted user data, the KRAKEN marketplace uses MPC with the SCALE-MAMBA protocol~\cite{scalemamba}.
Since MPC distributes the computation on multiple nodes ($N > 1$), a data seller needs to split their data into $N$ parts (shares) before encrypting and publishing it.
That is done using a Shamir secret-sharing-based protocol~\cite{DBLP:conf/scn/KellerRSW18}, which allows the nodes to use MPC to perform computations on the split data.
The protocol also guarantees that no node ever learns the complete data they are computing on.
After the computation, each node only possesses a share of the computation result, which the buyer can combine into the final result.
No one but the buyer may combine the result shares since they are encrypted using the buyer's public key by the nodes before returning them.

\mypar{Policy System}
We add our policy interpreter component to all nodes of the computation system (see \Cref{sec:securityassumptions} for why this is necessary).
As a policy system, we use an extended version of the \textit{TPL} system from the LIGHTest project (cf.~\Cref{sec:background:tpl})~\cite{DBLP:conf/ifiptm/ModersheimSWMA19, DBLP:conf/openidentity/AlberMMS21}.
TPL uses a syntax similar to Prolog and allows for expressive policies, which data sellers can use to restrict who can buy their data.
Most importantly, data sellers can use TPL to specify restrictions who can buy the data and what computations a buyer can perform.
Additionally, to establish trust in the buyer's attributes, TPL supports both the verification of certificates from more traditional trust schemes like Europe's eIDAS as well as self-sovereign identity (SSI) verifiable credentials.

The TPL interpreter is provided as a Java microservice, which we run on the nodes besides the computation service as described in \Cref{sec:arch}.

\mypar{Seller Policy in TPL}
To illustrate the syntax and structure of a TPL policy, in \Cref{fig:tpl} we give an example seller's policy in TPL\@.
The TPL interpreter is always started with a query for the \textit{accept} predicate.
In our KRAKEN TPL implementation, we define this predicate with three parameters:
The buyer's credentials are needed to check if a buyer is qualified to perform a computation.
In the example, the credentials are contained in a verifiable presentation, which is a structure common in the context of SSI~\cite{w3cVC}.
The other parameters are the number of records the buyer selected, and the computation type, which is extracted from the buyer's computation request by the node.

The given policy defines which credential data formats the interpreter's parser can expect (\textit{set\_format} predicate), and that it only accepts credentials that are qualified in the eIDAS trust scheme (\textit{check\_eIDAS\_qualified} predicate).

The data seller then defines that they only allow computations on their data if the data buyer selected at least 100 records.
This is done for privacy reasons, i.e., it is another mechanism to ensure that a computation does not reveal the seller's personal data in plaintext.
Also, the seller defined in their policy which type of buyer can initiate which type of computation on their data:
In our example, buyers from public universities can train machine learning models on the data, while private research institutions may only compute simple statistics.
Since the \emph{acceptComputaiton} predicate exists twice, and only one of them needs to yield \textit{true}, this construct constitutes a logical \textit{or}-operator.

\begin{figure}[ht]
  \begin{lstlisting}[language=Prolog, basicstyle=\ttfamily\footnotesize]
accept(BuyerCreds, NumRecords, ComputationType) :-
  set_format(BuyerCreds, w3c_verifiablePresentation),
  extract(BuyerCreds, mainCredential, BuyerCredential),
  set_format(BuyerCredential, w3c_verifiableCredential),

  extract(BuyerCredential, issuer, Issuer),
  check_eIDAS_qualified(Issuer),

  NumRecords > 100,
  extract(BuyerCredential, organization_type, OrgType),
  acceptComputation(OrgType, ComputationType).

acceptComputation(OrgType, ComputationType) :-
  OrgType == public_university,
  ComputationType == machine_learning.

acceptComputation(OrgType, ComputationType) :-
  OrgType == private_research,
  ComputationType == simple_statistics.
  \end{lstlisting}
  \caption[]{\label{fig:tpl} Illustrative TPL policy formulated by a data seller (with signature and trust chain verification omitted).}
\end{figure}

\mypar{Aggregated Policies}
Assuming a scenario in which a data buyer wants to execute a single computation over multiple data products, those products are likely to come with different policies (since a different data seller provides each).
To buy some computation on that data, the buyer needs to fulfill \textit{all} of those policies.
Since the marketplace needs to know which credentials it needs to request from the buyer, it first precomputes the list of credentials that are required for those policies.
Thus, we support the aggregation of several policies, finally containing all the rules of all policies involved in the computation request.
An \textit{aggregated policy} is the logical conjunction of all policies, which we create by combining the entry-point predicates of all policies with a logical \textit{and}-operator.

\section{Discussion}
\label{sec:discussion}

In this section, we present the results of the performance evaluation of our reference implementation.
We also discuss the security/trust assumption and adversaries we considered, including the possible attacks we mitigate.

\subsection{Evaluation}
\label{sec:evaluation}

To evaluate the practicability of our implementation, we conducted a performance analysis of our implementation.
We measure the impact of the additional TPL interpreter Java component on the time a buyer needs to wait for a result.

As a baseline, we take the performance of the MPC system used by the KRAKEN marketplace, which utilizes the MPC protocol SCALE-MAMBA~\cite{scalemamba}.
Computations can take from a few seconds (for simple statistics) over a couple of hours (for training a simple machine learning model) to even longer for more complex computations or larger datasets.\footnote{The KRAKEN MPC system source code and benchmarks~\cite{KRAKEN4.3} were provided to us in private by the authors.}
In contrast, our approach introduces an additional overhead from one to ten seconds for realistic policies.

The overhead consists of two parts:
On the one hand, executing a TPL policy introduces additional network round-trips.
Depending on the policy, the interpreter may establish network connections to retrieve the trust status information.
The incurring delay depends on the network performance between the interpreter and the trust status registries.
As a reference, we measured the latency of some common network actions in trust policies.
We used the TPL interpreter's HTTPS client in our office network.
Loading the eIDAS root trust status list XML\footnote{https://ec.europa.eu/tools/lotl/eu-lotl.xml} took us 0.3~s ($\pm$~0.209~s), while resolving an identifier from the Ethereum ledger\footnote{we use the non-production Universal Resolver at https://dev.uniresolver.io} takes 0.471~s ($\pm$~0.302~s).
We note that several of those network lookups are identical for many policies.
Thus, the nodes should be able to cache the results.
For example, the TPL interpreter, per default, downloads the eIDAS trust status lists of all EU member states during initialization, so no additional network access is required for any further eIDAS trust scheme check.

On the other hand, we measured the time for the interpreter to load and evaluate a policy.
Apart from the trust status information the interpreter loads from the Internet, the performance of a policy execution depends on the complexity of a policy.
For benchmarking, we used the Java Microbenchmark Harness (JMH)\footnote{https://github.com/openjdk/jmh} in version 1.35 and OpenJDK 16.
We executed the benchmarks on the TPL reference implementation using a business notebook from 2018 with an Intel i7-8550U quad-core CPU running Ubuntu 21.10.
The results we present in \Cref{tab:evaluation} show the execution of typical policies.
We observed that the run-time grows linearly with the amount of executed policies.
Thus for computations involving many data products, the nodes must aggregate the policies before execution to improve the run-time.

Since the measured timings are neglectable compared to the latency of a typical MPC computation, we argue that the performance overhead is acceptable.
We note that we use the TPL reference implementation, which is not optimized for performance.
Thus, an instantiation of our proposed architecture with an optimized policy system may further increase practicability.

\begin{table}[t]
\centering
\caption{Runtime benchmarks of the TPL interpreter.}\label{tab:evaluation}
\begin{tabular}{llcr}
\toprule
\# policies & \# predicates/policy & \quad & run-time [s/op] \\
\cmidrule{1-2} \cmidrule{4-4}
1           & 3                    &       & 0.08  (± 0.03)  \\
1           & 20                   &       & 0.09  (± 0.03)  \\
1           & 100                  &       & 0.12  (± 0.07)  \\
\cmidrule{1-2} \cmidrule{4-4}
100         & 3                    &       & 7.68  (± 1.68)  \\
100         & 20                   &       & 8.18  (± 0.59)  \\
100         & 100                  &       & 10.16 (± 1.84)  \\
\bottomrule
\end{tabular}
\end{table}

\subsection{Security Assumptions}
\label{sec:securityassumptions}

The goal of our approach is that a buyer who does not fulfill the given policy for some data must never be able to launch a computation on this data.
Also, the marketplace must neither be able to launch computations on its own nor learn the computation results. %

For MPC, the KRAKEN system uses a \textit{fully-malicious protocol}, which assumes that from the set of $N$ MPC nodes, at least one is honest~\cite{DBLP:conf/primelife/KochKPR20}.
Given this assumption, no node must gain access to the plaintext or the computation result.

The trust in our system equals the combined trust in the set of all nodes which perform the MPC computation.
Since an honest node would never start the MPC computation if the respective buyer does not fulfill a given policy, this effectively prevents the other nodes from computing anything on the data.
Thus, if the stated MPC assumption holds, then the goal is achieved, and our approach can be considered secure under the same assumptions as an MPC system.

\subsection{Attacks}

The goal is to prevent any illegitimate entity from accessing personal data, launching a computation, or accessing the result of a computation.
As adversaries, we consider
\begin{enumerate*}
\item a buyer who does not fulfill the policy for a data product,
\item a malicious marketplace, and
\item a malicious computation node.
\end{enumerate*}
We consider the following attacks:

\mypar{A marketplace wants to access data, launch a computation, or access the result of a computation}
In private data marketplaces, the marketplace platform itself has, by definition, no access to the data since the data is only shared with the platform in encrypted form.
The privacy of the computation itself depends on the marketplace's architecture.
In our KRAKEN-based implementation, a marketplace can not launch a computation by itself.
Since all MPC nodes are needed to perform a computation, the marketplace would have to convince all the nodes that it fulfills the policy.
Additionally, the marketplace has no access to the computation result because the MPC nodes encrypt the result (shares) only for the buyer.
In some other architectures~\cite{DBLP:conf/icdcs/KoutsosPCT020}, a curious marketplace can freely launch computations on the data at will.
At the same time, it can view the result of a computation launched by a legitimate buyer without much effort.
Thus, such architectures require a marketplace to be trustworthy to some degree.

\mypar{An adversary replaces the seller's policy with a policy they fulfill}
The policies are stored at the marketplace and sent to the computation system in plaintext, together with the links to the (encrypted) data.
Thus, any adversary, e.g., a curious marketplace, could send the links to data they are interested in alongside a fake policy that they can fulfill.
Such an attack is not possible in our design since we cryptographically link the policy to the corresponding data.
We let the user add a hash of the policy to the data product before they encrypt the data in an authenticated way.
If an attacker tries to replace the policy, the hashes do not match, and the computation node aborts the process.

\mypar{An adversary replaces the buyer's public key to access computation results}
In our MPC-based implementation, all result shares are encrypted with the buyer's public key.
This public key is sent as part of a credential to the computation system.
To gain access to the computation result, a malicious marketplace could try to replace the buyer's public key with their own.
Alternatively, they could add a credential with their public key to the computation request.
The request would then fulfill the policy using the buyer's real credentials, but the nodes would encrypt the result for the wrong public key.
We prevent this attack by including a check in the seller's policy that ensures that all credentials belong to the same identity.
Since the public key is extracted from one of the credentials, this ensures that only the legitimate buyer's public key is used to encrypt the computation result.

\subsection{Future Work}

A relevant scenario for private data marketplaces is the deletion of data products by a data seller.
The first step in this process is that the seller deletes their data packages from the cloud.
However, that is not enough since the marketplace or other actors with access could have cached it.
Even after the seller deleted their data, such an actor could use the computation system and a (valid) set of credentials to perform computations.
To mitigate this, we propose a revocation mechanism for data products, which the node checks during policy evaluation.
Since TPL policies can already be used to formulate generic revocation
checks for buyer
identities~\cite[Section 8.2]{w3cDID}
and credentials~\cite[Section 7.10]{w3cVC}, these checks could also be applied to the data itself.
Future work is needed to evaluate the integration of such a revocation mechanism into a private data market.

Additionally, when implemented, the same \textit{data revocation} construct can also be used by the buyer of some data.
For example, if a buyer executes the revocation-check policy periodically (or each time they use some data), they can be sure that usage of this data is still allowed.
That supports them in staying compliant with data protection regulations.

\subsection*{Conclusion}
\label{sec:conclusion}

Private data marketplaces support data buyers like research departments with discovering and performing computations on personal data.
While existing marketplaces already allow privacy-preserving computations on this data, a data seller cannot control who can buy their data and what computations shall be allowed.

We present an access control extension for private data marketplaces.
This extension enables data sellers to define expressive policies on their data usage.
The computation system then enforces those policies, ensuring the seller's control over their data.

Furthermore, we provide a proof of concept implementation for the KRAKEN marketplaces.
In this implementation, we demonstrate how to apply our approach to a distributed MPC-based marketplace.
We discuss our implementation's trust and security properties, arguing that it is secure under the same assumptions taken by the KRAKEN system.
By measuring that the overhead introduced by our policy component is neglectable when compared to standard MPC data analytics computations, we show that our approach is practical too.

\begin{acks}
  This work was supported by the
  \grantsponsor{EUH2020}{European Union's Horizon 2020 research and innovation programme}{https://ec.europa.eu/programmes/horizon2020/en} under grant agreements \textnumero~\grantnum{EUH2020}{871473} (KRAKEN) and \textnumero~\grantnum{EUH2020}{959072} (mGov4EU).
\end{acks}

\bibliographystyle{ACM-Reference-Format}
\bibliography{dblp,bib}

\end{document}